# Active Control of Probability Amplitudes in a Mesoscale System via Feedback-Induced Suppression of Dissipation and Noise

Chaitanya Gupta,[1*] Aldo Peña Perez,[1] Sean Fischer,[1] Stephen Weinreich,[1] Boris Murmann[1] and Roger T. Howe[1]

[1] *Department of Electrical Engineering, Stanford University, Stanford, CA 94305-4070*

Corresponding author: *cgupta2@stanford.edu

ABSTRACT: We introduce a classical potentiostatic feedback mechanism that attenuates the dissipation in a quantum system arising from coupling to the surrounding thermodynamic bath, preserving the inter-state interference in an electronic excitation transfer (EET) process. A three-terminal potentiostat device applies a low-noise voltage bias to the terminals of the EET system and reduces the physical coupling between the quantum system and its environment. We introduce a classical equivalent circuit to model the environment-coupled excitation transfer in an elementary two-state system. This model provides qualitative insight into how classical feedback action affects the transition probabilities between the states and selectively reduces the dissipative coupling for one of the vibronic energy levels of the transfer system. Furthermore, we show that negative feedback results in persistent spectral coherence between the energy level of the decoupled state and the vibronic levels of the complementary state, making the decoupled vibronic channel a probe for characterizing the vibronic structure of the complementary channel of the EET system.

## I. INTRODUCTION

Environment-induced decoherence of quantum states has been extensively investigated in experimental[1-4] and theoretical literature,[5-7] with the consensus that coupling to the degrees of freedom of a thermodynamic bath is sufficient to average out coherent interference in a quantum system. Environmental coupling is largely responsible for the onset of classical behavior in quantum systems. It has been identified as the cause for the extinction of 'glory' oscillations[8, 9] in the scattering of magnetic monopoles by charged molecules due to coupling with the rotational degrees of freedom in the molecular bath, as well as for the loss of information in superposed states within a quantum computer, again due to coupling with a surrounding thermodynamic bath with several degrees of freedom.[10-13] Therefore, preservation of the superposition of quantum probability amplitudes requires reduced interactions between the system and bath, or a reduction in the number of bath modes that can interact with the system, for the case when the quantum system is coupled to a large number of modes.[14]

A scheme for preserving interferences between states would enable new room temperature systems exhibiting quantum behavior that could be applied to sensing, computing and energy conversion. As an example, persistent quantum coherent interferences of exciton waves are thought to boost the efficiency of EET processes,[15-17] and by extension, the efficiency of an EET transport-mediated photosynthetic process.[18, 19] In this paper, we demonstrate how controlling the excitation of an EET charge transfer system with a classical electronic negative feedback loop can prolong the coherence lifetime for the participating electronic states. An environment-coupled molecular system that is comprised of a single level donor and acceptor species, 'dressed' by a collective of bath vibrational modes, is used to model the charge transfer process, which leads to an equivalent circuit model in which the dynamical variables describe wavefunction probability amplitudes. The impact of feedback on wavefunction probability amplitudes can then be described in terms of the dynamical variables of the circuit model.

## II. NEGATIVE FEEDBACK CONTROL OF INTERACTIONS WITH THE THERMALIZED ENVIRONMENT

### A. Classical Oscillator

The equations of motion for a one dimensional particle (system) coupled to a bath of damping vibrational modes are given by

$$\dot{x}(t) = \frac{p(t)}{m}; \quad \dot{p}(t) = -V'(x) - \int_0^t \frac{\bar{\gamma}(t-t')p(t')}{m}dt' + \bar{\xi}(t), \tag{1.1}$$

where $\bar{\gamma}$ is the damping kernel, and $m$, $k_B$, $T$ and $V(x)$ are the system mass, Boltzmann constant, bath temperature and conservative potential respectively. $\bar{\xi}(t)$ is a Gaussian function with zero mean and correlation given by $\langle \bar{\xi}(t)\bar{\xi}(0) \rangle = mk_B T \operatorname{Re} \bar{\gamma}(t)$ in the classical limit. Thus, the amplitude of thermal disturbance acting on the system is related to the dissipative force exerted by the environment, subject to the assumption that the thermal reservoir is large enough such that the bath vibrational modes continue to stay in equilibrium throughout their interaction with the system.[20] These equations of motion are derived from the Hamiltonian description of the system and the environment, in which the environment is modeled as a collection of non-interacting harmonic oscillators (h.o.) and the interaction between the system and the environment is bilinear in the environment h.o coordinates and the system coordinate [21-24]. Under the assumption that the environment coordinates and momenta values are sampled from an equilibrium Boltzmann distribution,[25] the fluctuation dissipation relationship can be shown to hold. The equivalent bath temperature as seen by the system is given by $\langle \bar{\xi}(t)\bar{\xi}(0)\rangle / m \operatorname{Re} \bar{\gamma}(t)$. In the Markovian limit, when the environment-system interaction is without memory, $\bar{\gamma}(t) = 2\eta \delta(t)$. The real part of the parameter $\eta$ would represent an effective viscosity in a mechanical system, or could be interpreted as a linear resistance in an oscillatory electrical circuit. (Figure 1A)

The system in question, whether quantum or classical, is stimulated by the randomized environment-induced thermal disturbances, which are balanced by the dissipative forces as it moves in the field described by the potential. We propose an electronic feedback-based mechanism for the bandwidth-limited control of these thermal disturbances and the related damping forces. The case of electrical oscillators is considered here for demonstration purposes, but the proposed mechanism could be extended to mechanical systems as well. Specifically, a scheme is presented whereby the system is decoupled from the physical reservoir with which it is in contact and coupled to another bath of pre-specified spectral density, ensuring control over the bath's effective 'temperature' and the damping experienced by the system. A sequence of cascaded amplifiers ($A_1$, $A_2$) is configured to deliver a high gain, corrective signal proportional to the difference between $V_{set}$ and $V_{ref}$ upon measurement of the reference voltage, $V_{ref}$ (Figure 1B). The measurement is performed with a buffer amplifier, $B_1$ that has a high impedance input to minimize leakage currents in the measurement. The physical reference electrode (R.E.) for probing the reservoir voltage, $V_{ref}$, is deemed ideally to have zero source impedance, as is the physical counter electrode (C.E.) that applies the corrective signal $V_{x'}$ to the system (Figure 1B). In addition, the circuit schematics in Figures 1A, 1B measure the current flowing through the system across dissipative elements $X_r$ and $X_m$, respectively, in response to the classical voltage excitation bias applied at the reference electrode. The system impedance is characterized as $X+jY$, where $X$ represents the dissipative part and $Y$ is the purely imaginary contribution.

The respective transimpedance responses for systems in Figures 1A and 1B are:

$$I_x = \frac{V_{set}}{X_r - jY_s} \quad (2.1)$$

$$I_{x'} = \frac{A_1 A_2 V_{set}}{X_m - jY_s + A_2(1+A_1) \cdot (-jY_s)} \quad (2.2)$$

where $Y_s$ is the resonant component of the system, given by $Y_s = (1/\omega C_1 - \omega L_1)$. $A_1, A_2$ are the gain functions of the respective amplifiers; the dissipative elements, $X_r$ and $X_m$ in (2.1, 2.2), have real and imaginary components obtained by averaging over the ensemble of vibrational modes[25]:

$$X_r(\omega) = -j \cdot \left[ PP \int d\Omega\, g(\Omega) \frac{c^2(\Omega)}{m\Omega^2} \frac{1}{\omega^2 - \Omega^2} \right] + 2\pi \frac{g(\omega) c^2(\omega)}{m\omega^2} \quad (2.3)$$

where the first term, which includes the principle part (symbol $PP$) of the integral over the complex plane, is representative of a resonant frequency shift and the real term is the dissipation experienced by the oscillating system. Thus, the response of the oscillatory system to thermal excitations is dictated by the ensemble-averaged lumped circuit representation of the environment-induced dissipation as well as by the ensemble-averaged 'dressing' down of the resonant frequency of the system, also by the environment vibrational modes. The application of high gain negative feedback cancels the dissipation and the dressing down of the resonance as observed in the *LTSpice IV* A. C. small-signal simulation in Figure 2. Commercial operational amplifier schematics were used to generate the simulation results in Figure 2.

The thermal disturbances induced by the reservoir on the system are measured at the R. E. node. These disturbances can be estimated and referred to the input source $V_{excitation}$ for the schematics in 1A and 1B, as is standard practice in noise analysis in electronic circuits.

$$\langle V_x^2 \rangle = \frac{|Y_s|^2}{|X_r - jY_s|^2} \langle V_{X_r}^2 \rangle \tag{3.1}$$

$$\langle V_{x'}^2 \rangle = \frac{|A_1 A_2|^2 |Y_s|^2}{|X_m - jY_s + A_2(1+A_1)(-jY_s)|^2} \langle V_{ref}^2 \rangle \tag{3.2}$$

The input-referred noise at the reference node $\langle V_{ref}^2 \rangle$ is obtained by superposing the input referred voltage noise from each source in the feedback loop and referring them to the input:

$$\langle V_{ref}^2 \rangle = \left( \frac{\langle V_{A_1}^2 \rangle}{\Delta f_{A_1}} + \frac{\langle V_{B_1}^2 \rangle}{\Delta f_{B_1}} \right) \Delta f_{ref} \left| \frac{A_1 A_2}{1+A_1 A_2} \right|^2 + \left( \frac{\langle V_{A_2}^2 \rangle}{\Delta f_{A_2}} + \frac{\langle V_{X_r}^2 \rangle}{\Delta f_{X_r}} + \frac{\langle V_{X_m}^2 \rangle}{\Delta f_{X_m}} \right) \Delta f_{ref} \frac{1}{|1+A_1 A_2|^2} \tag{3.3}$$

and $\Delta f_i$ is the bandwidth of the *i*-th voltage noise source. For large $X_m$ and $A_1$, small reference node bandwidth $\Delta f_{ref}$ and a sufficiently quiet feedback network, the system would experience significantly smaller thermal disturbances, or a lower equivalent bath temperature, than in the case without feedback. The equivalent mode temperature for the oscillatory system, when in equilibrium with the reservoir modes, is estimated from the equipartition theorem as $T_s = (1/2\pi C_s k_B) \cdot \int \langle q_s^2 \rangle d\omega$ [26] where

$$\langle q_s^2 \rangle = \frac{1/L_s^2}{\frac{X_m^2 \omega^2}{L_s^2 |A_1 A_2|^2} + (\omega_s^2 - \omega^2)^2} \frac{\langle V_{ref}^2 \rangle}{\Delta f_{ref}} \quad (3.4)$$

assuming $|A_1 A_2| \gg 1$ and is independent of frequency, and for which $\omega_s^2 = 1/L_s C_s$. Integrating over the frequency domain yields

$$T_s \sim T \cdot \left(1 + \frac{X_r}{X_m}\right) \frac{|A_1 A_2|}{|1 + A_1 A_2|^2} + \left(\frac{\langle V_{A_2}^2 \rangle}{\Delta f_{A_2} X_m k_B}\right) \frac{|A_1 A_2|}{|1 + A_1 A_2|^2} + \left(\frac{\langle V_{A_1}^2 \rangle}{\Delta f_{A_1}} + \frac{\langle V_{B_1}^2 \rangle}{\Delta f_{B_1}}\right) \frac{|A_1 A_2|}{X_m k_B} \left|\frac{A_1 A_2}{1 + A_1 A_2}\right|^2 \quad (3.5)$$

In effect, the physical environment around the system is exchanged with the bath of modes associated with the measurement and feedback instrumentation, which can be tailored for a lower equivalent bath temperature by choosing components $A_1$, $A_2$, $B_1$ and $X_m$ with minimal thermal noise characteristics. This method of electronic 'cooling' can be contrasted with other active feedback-based methodologies in opto-mechanical systems that utilize a large gain to *increase* the dissipative coupling between the mechanical system and its single mode optical environment, pre-prepared in a low temperature state, for improved cooling efficiency.[27-30] Simulations of voltage noise spectral density, $\langle V_x^2 \rangle / \Delta f$ and $\langle V_{x'}^2 \rangle / \Delta f$, as functions of $X_r$ are depicted in Figures 3A and 3B respectively. As the simulations indicate, feedback 'cools' the system, with the largest damping kernel being cooled the most. The reduction in total integrated noise power with increasing $X_m$, as observed from the reduced area under the curve, and the corresponding equivalent system mode temperature, is also illustrated in Figure 3C.

## B. Mesoscopic Charge Transfer System

The quantum dynamics of Hermitian Hamiltonians are known to correspond to the coupled motion of classical mechanical or electrical oscillators.[31, 32] Specifically, the classical probability amplitudes describing the time-dependent state of an oscillatory system are equivalent to the quantum amplitudes that characterize the evolution of the wavefunction of an excited quantum system by a time-dependent Schrodinger's wave equation.[31] Consider single energy level donor and acceptor states, immersed in a reservoir bath, and coupled to one another so as to excite an electronic transition from the electronic source to the sink.[25, 32]

$$
\begin{aligned}
H &= H_d + H_a + H_{env} + H_{d-a} + H_{d-env} + H_{a-env} \\
&= \frac{\bar{p}_1^2}{2M_1} + \frac{M_1 \omega_1^2 \bar{Q}_1^2}{2} + \frac{\bar{p}_2^2}{2M_2} + \frac{M_2 \omega_2^2 \bar{Q}_2^2}{2} + \sum_\alpha \frac{\bar{p}_\alpha^2}{2m_\alpha} + \frac{m_\alpha \omega_\alpha^2 \bar{q}_\alpha^2}{2} \\
&\quad + V_{12} \bar{Q}_1 \bar{Q}_2 + \sum_\alpha \left( V_{1\alpha} \bar{Q}_1 + V_{2\alpha} \bar{Q}_2 \right) \bar{q}_\alpha
\end{aligned}
\quad (4.1)
$$

The 'momentum' and 'position' coordinates for the system and environment can be suitably non-dimensionalized [31, 32] to re-derive the dynamics of the system and the environment from the modified non-dimensionalized Hamiltonian:

$$
\begin{aligned}
&\dot{Q}_i(t) = \omega_i p_i; \quad \dot{p}_i(t) = -\omega_i Q_i + v_{ik} Q_k + \sum_\alpha v_{i\alpha} q_\alpha \\
&i, k = 1, 2; \; k \neq i
\end{aligned}
\quad (4.2)
$$

$$
\dot{q}_\alpha(t) = \omega_\alpha p_\alpha; \quad \dot{p}_\alpha(t) = -\omega_\alpha q_\alpha + v_{1\alpha} Q_1 + v_{2\alpha} Q_2 \quad (4.3)
$$

Based on the dynamical equations of motion (4.1-4.3), we postulate an equivalent circuit description of the inter-coupled, single electronic energy level, donor and acceptor charge transfer system in Figure 4A. The individual energy levels are modeled as resonant elements that are coupled to the physical environment or, in the case of feedback, to the bath reservoir of instrumentation modes dissipatively via resistors at the reference node. The reference node in Figure 4A defines the location where an external bias is applied, or where, in the feedback case, a reference probe is inserted to measure the 'energy' of one of the levels that the feedback loop constrains to a desired setpoint. In this context, the reference is deemed a proxy measure for the second energy level in experimental systems where direct access to the state energy is unavailable. In addition, a capacitor between a resonant unit and the reference probe models the non-dissipative coupling between the energy levels. The 'ground' for the proposed circuit model in Figure 4A defines the energy ground state relative to which the energy of the resonant elements ($\omega_1$, $\omega_2$) and the source excitation signal ($eV/\hbar$) are measured. Recasting the equations of motion in terms of the probability amplitude for donor/acceptor states, $Z_{1,2} = Q_{1,2} + jp_{1,2}$, as well as for the environment modes, $Z_\alpha = q_\alpha + jp_\alpha$,

$$\dot{Z}_i + j\omega_i Z_i = jv_{ik}Z_k - \sum_\alpha v_{i\alpha}^2 \int_0^t dt' Z_i(t') e^{-j\omega_\alpha(t-t')}$$

$$- \sum_\alpha v_{i\alpha}v_{k\alpha} \int_0^t d\tau Z_k(\tau) e^{-j\omega_\alpha(t-\tau)} + j\sum_\alpha v_{i\alpha} Z_{\alpha 0} e^{-j\omega_\alpha t} \quad (4.4)$$

$$i, k = 1, 2; i \neq k$$

after integrating out the environment mode dynamics. The state occupation probabilities may be estimated from $P_i = \dfrac{|Z_i|^2}{\sum_i |Z_i|^2}$. $Z_{\alpha 0}$ is the randomly chosen initial value for the occupation probability of mode $\alpha$. The form of the dynamical equation (4.4) constitutes a Hermitian generalization of the Hamiltonian in (4.1) with linear position and momentum off-diagonal coupling, which is also referred to as a system of p&q coupled oscillators.[32] The last terms on the R.H.S. of equation 4.4 constitute the noise source terms that thermally excite the transfer events. Equation 4.4 can be transformed by a redefinition of the variables $Z_{1,2}e^{j\omega_{1,2}t} = \tilde{Z}_{1,2}$ resulting in

$$\dot{Z}_i = -\left(j\omega_i + j\Delta\omega_{ii} + \frac{1}{2}\gamma_{ii}\right)Z_i + \left(jv_{ik} - j\Delta\omega_{ik} - \frac{1}{2}\gamma_{ik}\right)Z_k + j\sum_\alpha v_{i\alpha}Z_{\alpha 0}e^{-j\omega_\alpha t} \quad (4.5)$$

$$i,k = 1,2; i \neq k$$

with the parameters $\Delta\omega_{mn} = PP\int d\omega g(\omega)\frac{v_m(\omega)v_n(\omega)}{\omega_n - \omega}$ and $\gamma_{mn} = 2\pi v_m(\omega_n)v_n(\omega_n)g(\omega_n)$.

Specific cases for large and small inter-level coupling ($v_{12}$) are considered as asymptotic limits of the proposed 'classical' charge transfer model. The model is simplified by the assumption that bath modes for the two charge transfer component systems are identical, *i.e.* $v_{1\alpha} = v_{2\alpha} = v_\alpha$ for all $\alpha$, without any loss in generality. For the case when $v_{12} \gg \max\left(|\omega_1 - \omega_2|, \left|\int_0^\infty d\tau \sum_\alpha v_\alpha^2 e^{-j\omega_\alpha\tau}\right|\right)$, the eigenvalues of the system (4.5) are

$$\Omega_1 = \frac{\omega_1 + \omega_2}{2} - v_{12} + (\Delta\omega_{11} + \Delta\omega_{22}) - j\frac{(\gamma_{11} + \gamma_{22})}{2} \; ; \; \Omega_2 = \frac{\omega_1 + \omega_2}{2} + v_{12} \quad (4.6a)$$

In the limit that $\Delta\omega_{ii}, \gamma_{ii} \to 0$ for $i = 1,2$, the results in (4.6) indicate the creation of new energy surfaces engendered by a split of magnitude $2v_{12}$ in the strongly coupled h.o. wells of energy $\omega_1$ and $\omega_2$. The occupation probabilities for these new energy surfaces, as functions of time, are given by:

$$P(\omega = \Omega_1, t : \Delta\omega_{ii}, \gamma_{ii} \to 0) = 1 \; ; \; P(\omega = \Omega_2, t : \Delta\omega_{ii}, \gamma_{ii} \to 0) = 0 \quad (4.6b)$$

for the case when the initial condition requires that the system in populated in state $\omega_1$. These results are indicative of an adiabatic transfer process, characterized by a confinement of the electronic charge to an adiabatic energy surface through the process of transfer from the donor to the acceptor.[33] On the other hand, when $v_{12} \to 0$, the eigenvalues are given by

$$\Omega_1 = \omega_1 + \Delta\omega_{11} - j\frac{\gamma_{11}}{2}; \quad \Omega_2 = \omega_2 + \Delta\omega_{22} - j\frac{\gamma_{22}}{2} \tag{4.7a}$$

which is indicative of a diabatic crossing[25] of the weakly coupled h.o. wells, also for the limit of zero dissipation. The corresponding probability that the system makes a quantum jump from energy surface $\omega_1$ to surface $\omega_2$ at the crossing of the diabatic curves is given by:

$$P(\omega = \Omega_2, t : \Delta\omega_{ii}, \gamma_{ii} \to 0) = v_{12}^2 \frac{\sin^2\left(\frac{(\omega_1 - \omega_2)}{2}\right)t}{\left(\frac{\omega_2 - \omega_1}{2}\right)^2} \quad \text{and}$$

$$P(\omega = \Omega_1, t : \Delta\omega_{ii}, \gamma_{ii} \to 0) = 1 - P(\omega = \Omega_2, t : \Delta\omega_{ii}, \gamma_{ii} \to 0) \tag{4.7b}$$

in the limit of vanishingly small $v_{12}$ and by ignoring the effects of the environment on the transition process. The transition probability as derived from (4.7b) obeys Fermi's golden rule.[25] The results in (4.6-4.7) confirm the applicability of the classical model in describing charge transfer in the asymptotic limits of adiabatic and diabatic transfer. The inclusion of the effects of the reservoir bath in the estimation of the eigenfrequencies and of the transition rate for the diabatic case indicates that the excitation due to the coupling between donor and acceptor states can be dissipated through the many mechanisms for energy-exchange that exist between the charge transfer system and the external reservoir. The transition probability for the electron to make the jump from energy surface $\Omega_1$ to $\Omega_2$ is determined by indirect paths through the environment by the inelastic exchange of energy between the individual states $\Omega_1$, $\Omega_2$ and the reservoir modes that are determined by the density of states of the environmental modes and their coupling strength to the charge transfer system, which contribute to the dissipation experienced by the two-level system. The resulting environment-induced damping is responsible for the rapid extinguishing of coherent interference between the two states. The application of an electronic feedback mechanism to attenuate the damping induced by the physical reservoir would, as we will show, (a) minimize the non-dissipative coupling between the two energy states, rendering the EET process diabatic, and (b) enable the preservation of the coherent interference phenomena between vibronic states of the two-level systems. A simultaneous reduction in the r.m.s voltage fluctuations between the participant energy states by a low voltage-noise feedback mechanism would also help suppress the background due to the inelastic processes.

The two-state charge transfer system, coupled to the electronic feedback loop, is depicted in Figure 4B. Here, the reference probe measures the energy of the quantum state 2, and the feedback sets the energy to a desired setpoint via a corrective signal applied to the energy of state 1. All state energies are measured relative to the system ground as mentioned previously. We shall assume an ideal, dissipation-free reference probe in contact with the participant energy state 2 for the subsequent analysis with $\Delta\omega_{22}, \gamma_{22} \to 0$. The analysis may be extended to the more general case with dissipation in the reference channel. With the application of feedback, the probability amplitudes for the states of the two participatory species in the charge transfer process are given by:

$$Z_1(V,\omega) = \frac{\left(-jv_{12}\dfrac{j\Delta\omega_{11}+\gamma_{11}/2}{(jv_{12}+j\Delta\omega_{11}+\gamma_{11}/2)}+j\left(\dfrac{eV}{\hbar}-\omega_2\right)\right)\sum_\alpha v_\alpha Z_{\alpha 0}\delta\left(\dfrac{eV}{\hbar}-\omega_\alpha\right)}{\left(j\Delta\omega_{11}+\gamma_{11}/2+j\left(\dfrac{eV}{\hbar}-\omega_1\right)\right)\left(-jv_{12}\dfrac{(j\Delta\omega_{11}+\gamma_{11}/2)}{(jv_{12}+j\Delta\omega_{11}+\gamma_{11}/2)}+j\left(\dfrac{eV}{\hbar}-\omega_2\right)\right)\dfrac{1}{A_1 A_2(\omega)}+j\left(\dfrac{eV}{\hbar}-\omega_1\right)\cdot j\left(\dfrac{eV}{\hbar}-\omega_2\right)}$$

(5.1a)

$$Z_2(V,\omega) = \frac{j\left(\dfrac{eV}{\hbar}-\omega_1\right)\sum_\alpha v_\alpha Z_{\alpha 0}\delta\left(\dfrac{eV}{\hbar}-\omega_\alpha\right)}{\left(j\Delta\omega_{11}+\gamma_{11}/2+j\left(\dfrac{eV}{\hbar}-\omega_1\right)\right)\left(-jv_{12}\dfrac{(j\Delta\omega_{11}+\gamma_{11}/2)}{(jv_{12}+j\Delta\omega_{11}+\gamma_{11}/2)}+j\left(\dfrac{eV}{\hbar}-\omega_2\right)\right)\dfrac{1}{A_1 A_2(\omega)}+j\left(\dfrac{eV}{\hbar}-\omega_1\right)\cdot j\left(\dfrac{eV}{\hbar}-\omega_2\right)}$$

(5.1b)

In these descriptors for the probability amplitudes, the excitation signal applied to the EET system via the feedback loop input, seen in Figure 4B, comprises two separable frequency components: a high frequency part that characterizes the energy difference between the two participant states ($V$) of the quantum mechanical charge transfer system and a low frequency signal that describes the time response of the electrical feedback mechanism ($\omega$). In the asymptotic limit of large gain, the dynamic equations, as derived from Equations (5.1a) and (5.1b), governing the evolution of the probability amplitudes are given by:

$$\dot{Z}_1 = -j\omega_1 Z_1 - j(\omega_2-\omega_1)\frac{jv_{12}(j\Delta\omega_{11}+\gamma_{11}/2)}{jv_{12}+j\Delta\omega_{11}+\gamma_{11}/2}Z_2; \quad Z_1(0) = v(\omega_1)g(\omega_1)Z_{10} \quad (5.2a)$$

$$\dot{Z}_2 = -j\omega_2 Z_2; \quad Z_2(0) = v(\omega_2)g(\omega_2)Z_{20} \quad (5.2b)$$

where the significantly slower dynamics of timescales $\sim 1/\omega$ are considered static as the probability amplitudes rapidly evolve towards steady state. The eigenfrequencies for the charge transfer system, in the limit of large gain, and for the specific case of the dissipation-less reference probe are given by:

$$\Omega_1(\Delta\omega_{22}, \gamma_{22} \to 0) = \omega_1 \, ; \, \Omega_2(\Delta\omega_{22}, \gamma_{22} \to 0) = \omega_2 \qquad (5.3)$$

which we note are *independent* of the non-dissipative coupling $v_{12}$ between the participant species. The feedback decouples the interacting energy states from one another and constrains the EET process to be diabatic in nature. Therefore, a linear sweep of the voltage at the reference node, where $\omega_2 = \omega_2^o - eV/\hbar$, is analogous to a scan of the energy of state 2. The r.m.s. voltage noise determines the spread around the frequency $\omega_2$ and a low-noise voltage excitation signal mitigates this spread, which is analogous to the effect of a cryostatic reduction in bath temperature, as we now demonstrate.

The participating species in the transfer process are indistinguishable from the environment at t=0 and the probability amplitudes of environment modes of frequencies $\omega_1$, $\omega_2$ are $Z_{10}$ and $Z_{20}$ respectively. The environment modes are assumed to evolve along a deterministic trajectory determined by the dynamics of the classical excitation signal, $V$, acting on the modes. As such, amplitudes of environment modes at energies $\omega_1$ and $\omega_2$ are described by their respective coherent state amplitudes[34, 35] as

$$Z_{i0}(\omega) = \frac{1}{(e\Delta V/\hbar)^{1/2} \pi^{1/4}} \exp\left(-\frac{(\omega-\omega_i)^2}{2(e\Delta V/\hbar)^2}\right); \quad i=1,2 \qquad (5.4)$$

Here $\Delta V$ is the thermal r.m.s. voltage fluctuation of the excitation signal, which is proportional to $\sqrt{T}$.[36] The corresponding initial conditions in Equations (5.2a) and (5.2b) would be modified as

$$Z_1(0) = \int_{-\infty}^{\infty} v(\omega) g(\omega) Z_{10}(\omega) \, ; \, Z_2(0) = \int_{-\infty}^{\infty} v(\omega) g(\omega) Z_{20}(\omega) \qquad (5.5)$$

The spread about the environment mode frequencies $\omega_1$ and $\omega_2$, $e\Delta V/\hbar$, determines whether bath modes in the vicinity of the characteristic frequencies are able to contribute to the evolution of the wavefunctions for the sub-systems 1 and 2 that are participating in the EET process. Minimization of the r.m.s voltage noise at the reference node of the feedback loop or an equivalent reduction in bath temperature reduces the contribution from these background processes for states 1 and 2. Thus, environment-induced scattering into and out of the electronic states 1 and 2 is confined to bath modes that are resonant with the state energies $\omega_1$ and $\omega_2$.

The solution of the dynamical equations (5.2a) and (5.2b) yield the time evolution of probability amplitudes for states 1 and 2:

$$Z_1(t) = v(\omega_1) g(\omega_1) Z_{10} \left( 1 - \frac{jv_{12}(j\Delta\omega_{11} + \gamma_{11}/2)}{(jv_{12} + j\Delta\omega_{11} + \gamma_{11}/2)} \right) e^{-j\omega_1 t}$$
$$+ v(\omega_2) g(\omega_2) Z_{20} \left( \frac{jv_{12}(j\Delta\omega_{11} + \gamma_{11}/2)}{(jv_{12} + j\Delta\omega_{11} + \gamma_{11}/2)} \right) e^{-j\omega_2 t}$$ (5.6a)

$$Z_2(t) = v(\omega_2) g(\omega_2) Z_{20} e^{-j\omega_2 t}$$ (5.6b)

for the ideal initial conditions of zero spread about the environment modes $\omega_1$ and $\omega_2$. The line width around the *electronic state* $\omega_1$ is also minimized by the attenuation of the dissipative coupling between state 1 and its environment modes which has been described previously in equation (2.2). Therefore, the primary EET process is constrained to an exchange of energy between the electronic energy level of state 1 and the bath mode at frequency $\omega_2$, where each participant state energy level is characterized by a narrow spread. The participant electronic states also exchange energy with bath modes that are resonant with the respective electronic energies. State 1, for which the feedback attenuates the dissipative coupling with the environment modes, is also characterized by persistent spectral coherence with the bath mode resonant with state 2 as seen in Equation (5.6a). The interference between the electronic and vibronic states, observed within the dynamic variables $Q_1, p_1$ that characterize an EET participant, enables measurement of the vibronic structure of the complementary participant that is subject to the energy scan. This measurement methodology is particularly useful where direct measurement of the dynamic variables of the complimentary participant in the EET process is not possible, for example in a molecular electrochemical charge transfer system, where state 2 characterizes a redox-active molecule dissolved in a liquid electrolyte medium.

## III. CONCLUSIONS

In summary, we have proposed a three-terminal negative feedback mechanism that attenuates dissipation from a thermodynamic bath to preserve coherent interferences between participant states in an EET process. A classical circuit analogy is shown to characterize the effect of electronic feedback on the quantum EET system. In addition, the dissipation-free state can probe the vibronic characteristics of the complementary participant state through the suppression of the r.m.s. voltage fluctuations between the two states using negative feedback.

## IV. ACKNOWLEDGEMENTS

This work was supported by the Defense Advanced Research Projects Agency grant N66001-11-1-4111, through the Mesodynamic Architectures program, Dr. Jeffrey Rogers, Program Manager.

**Figure 1**

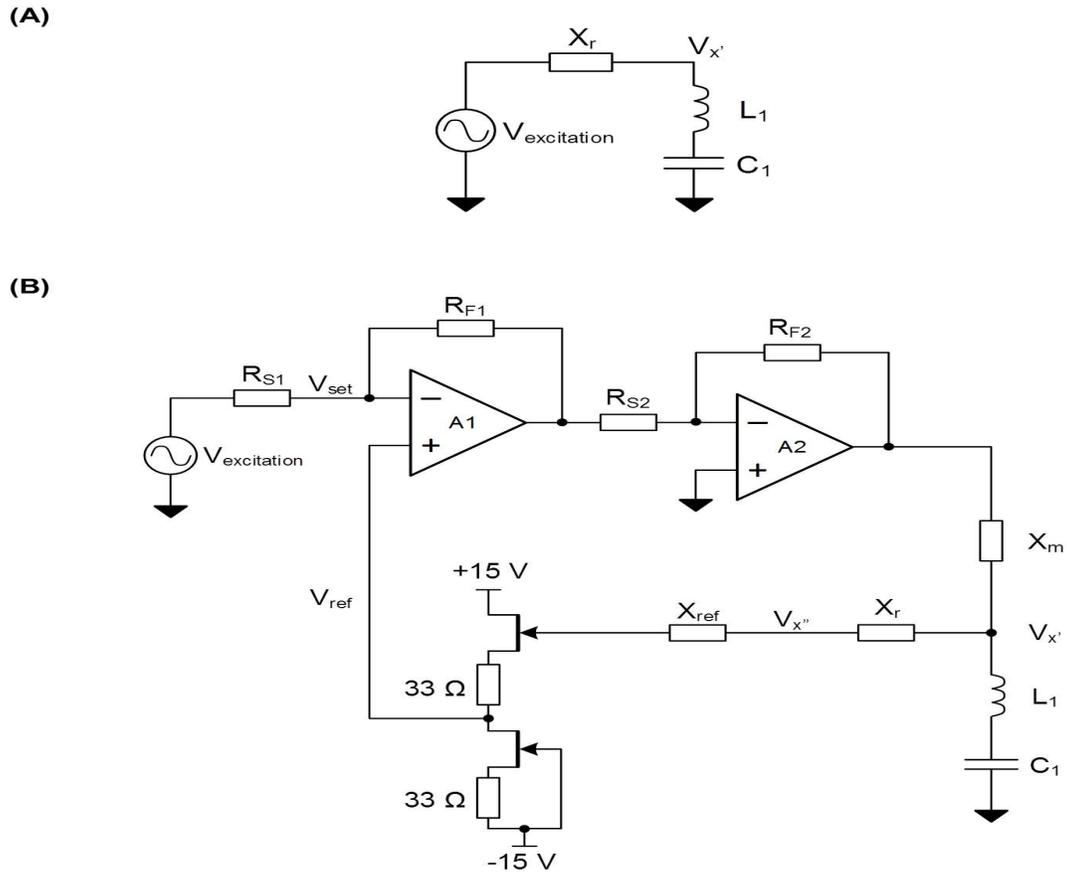

FIGURE 1. (A) Schematic illustrating the circuit model for an oscillator (characterized by L1, C1) in contact with a thermodynamic bath, with a classical excitation source applying a signal to the oscillator system via a dissipative contact. (B) Three terminal (W.E., R.E. and C.E.) feedback system for the application of the same signal to the oscillatory system, where gain in the feedback loop attenuates the dissipative coupling to the environment. Nodes x and x' are marked in (A) and (B) respectively.

**Figure 2**

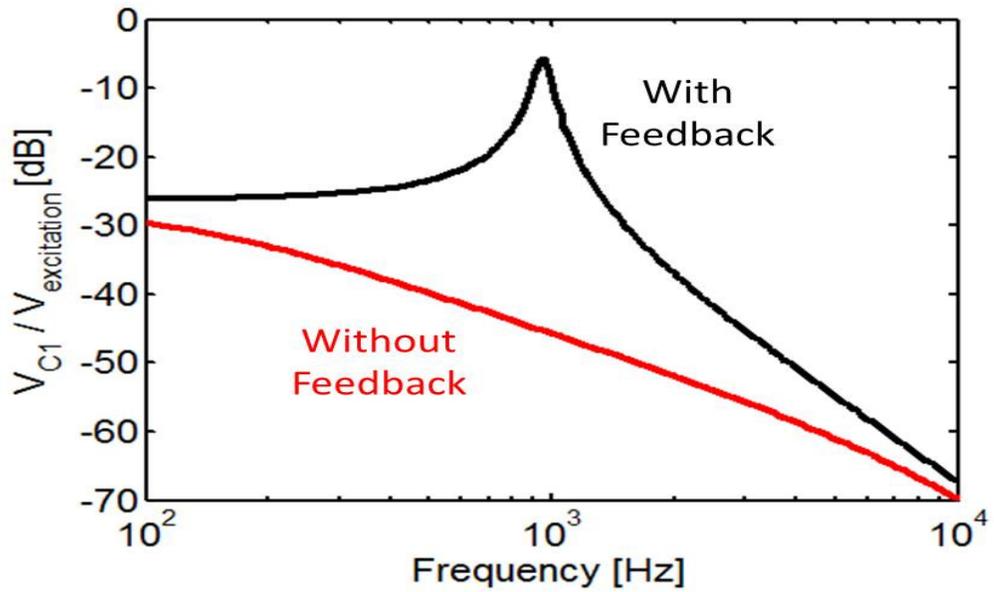

FIGURE 2. Graph for the oscillation amplitude measured at C1, in response to a small signal AC excitation~0.4V, with and without feedback. For the *LTSpice IV* simulation, amplifiers were selected from its component library. L1=0.198H, C1=142nF, $X_r$=9878ohm and $X_m$=100kohm for these calculations.

**Figure 3**

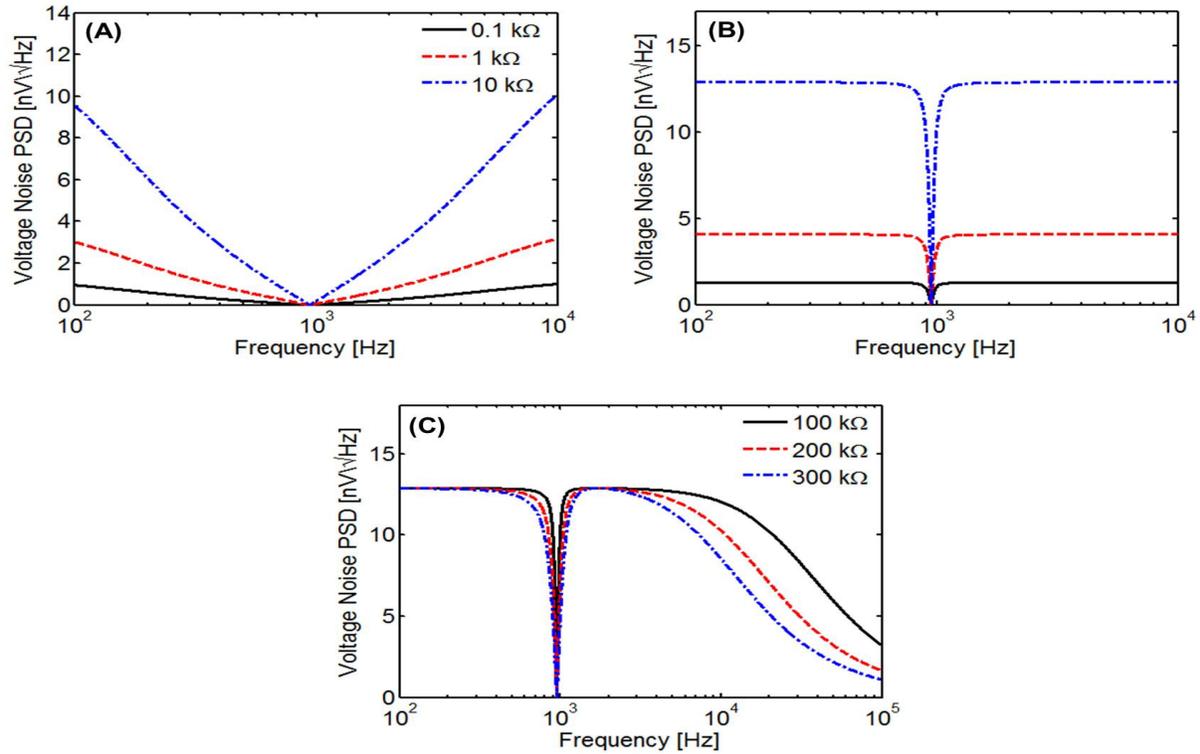

FIGURE 3. (A) Spectral density of voltage noise for the oscillator node x, without feedback as a function of $X_r$. (B) Spectral density of voltage noise for the oscillator node x', with feedback, as a function of $X_r$. The series LC construct, representing the oscillator system, creates a high Q bandpass filter at node x', as a result of dissipation attenuation by the feedback. (C) Spectral density of voltage noise for the oscillator node x', with feedback, as a function of $X_m$. Larger $X_m$ result in a lower total voltage noise power, yielding a lower effective bath temperature.

**Figure 4**

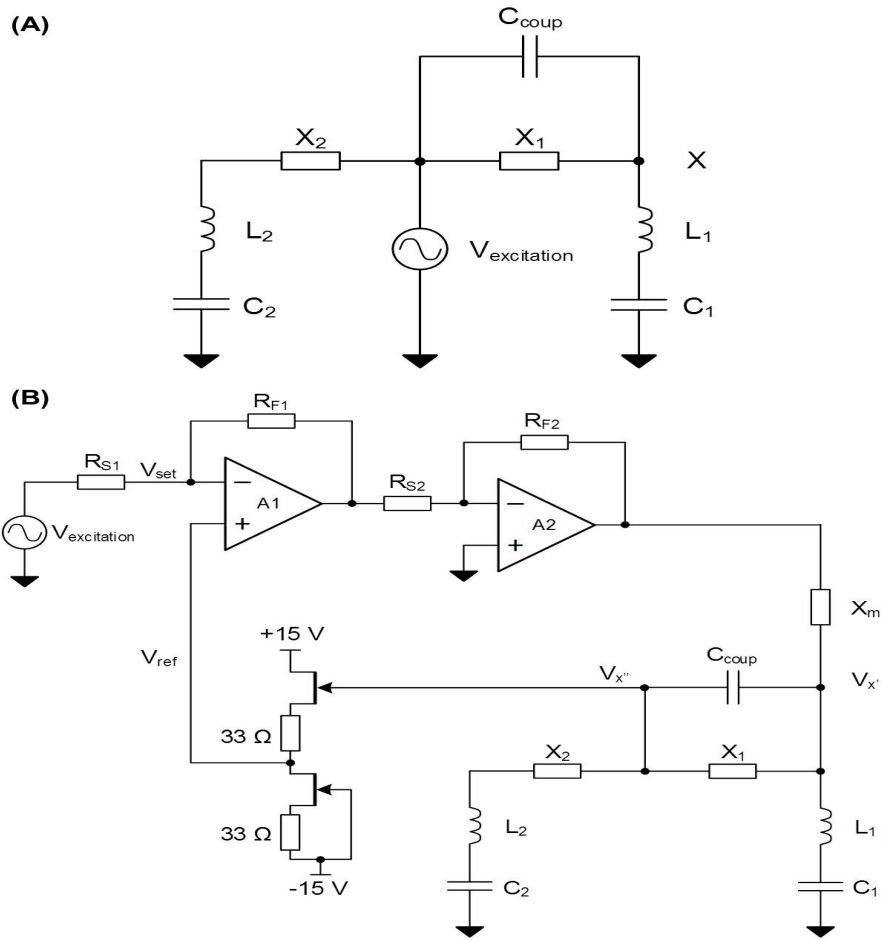

FIGURE 4. (A) Schematic illustrating circuit equivalent for a two state EET charge transfer system that is coupled to external bath of reservoir modes. (B) Feedback coupled to EET system for the attenuation of environment-induced dissipation